\documentclass{aes2e}
\usepackage{url}
\usepackage[hidelinks]{hyperref}
\usepackage{doi}

\jyear{2022}
\jmonth{September}
\jvol{70}
\jnum{9}

\begin{document}
\markboth{Venkatesh, Moffat and Miranda}{Word Embeddings for Automatic Equalization in Audio Mixing}

\title{Word Embeddings for Automatic Equalization in Audio Mixing\thanks{To whom correspondence should be addressed e-mail: satvik.venkatesh@plymouth.ac.uk}}

\authorgroup{
	\author{SATVIK VENKATESH$^1$},
	\author{DAVID MOFFAT$^2$},
	\role{AES Member}
	AND \author{EDUARDO RECK MIRANDA$^1$}
	\email{(satvik.venkatesh@plymouth.ac.uk) (dmof@pml.ac.uk) \quad\quad\quad\quad\quad\quad\quad\quad\quad\quad\quad\quad (eduardo.miranda@plymouth.ac.uk)\quad\quad}
	\affil{$^1$ Interdisciplinary Centre for Computer Music Research, University of Plymouth, Plymouth, UK \\ $^2$ Plymouth Marine Laboratory, Plymouth, UK}
}

\abstract{%
In recent years, machine learning has been widely adopted to automate the audio mixing process. Automatic mixing systems have been applied to various audio effects such as gain-adjustment, equalization, and reverberation. These systems can be controlled through visual interfaces, providing audio examples, using knobs, and semantic descriptors. Using semantic descriptors or textual information to control these systems is an effective way for artists to communicate their creative goals. In this paper, we explore the novel idea of using word embeddings to represent semantic descriptors. Word embeddings are generally obtained by training neural networks on large corpora of written text. These embeddings serve as the input layer of the neural network to create a translation from words to EQ settings. Using this technique, the machine learning model can also generate EQ settings for semantic descriptors that it has not seen before. We compare the EQ settings of humans with the predictions of the neural network to evaluate the quality of predictions. The results showed that the embedding layer enables the neural network to understand semantic descriptors. We observed that the models with embedding layers perform better than those without embedding layers, but still not as good as human labels. 
}

\maketitle
\setcounter{section}{0}
\section{INTRODUCTION}
The process of audio production involves multiple tasks such as balancing sound levels and applying audio effects. An audio effect can be defined as a function that transforms sound based on a set of controlled parameters~\cite{wilmering2020history}. Audio production is needed in various domains such as making albums, films, and theatre works, to name a few. It is generally carried out by a mixing engineer who understands the goals of their client. The mixing engineer blends multiple tracks together by modifying acoustic properties such as dynamics and timbre~\cite{ramirez2021deep}. A vast body of research has been exploring how this process can be automated through the use of \emph{intelligent tools}~\cite{de2013knowledge, de2019intelligentmusic, moffat2018towards, perez2009automatic}. Traditional Artificial Intelligence (AI) approaches such as expert systems have been adopted to create autonomous mixing tools~\cite{de2013knowledge}. These systems are knowledge-engineered and adopt a set of rules for mixing depending on the scenario. However, recent research has grown towards using Machine Learning and Deep Learning for automatic mixing. On one hand, some studies have focused on specific areas, such as gain balancing~\cite{moffat2019machine} and reverberation~\cite{chourdakis2017machine}. On the other hand, some have explored building autonomous systems where the entire mixing process is carried out without human intervention~\cite{ramirez2021deep, moffat2019approaches}. 

An equalizer (EQ) is an audio effect created by cascading multiple filters in series~\cite{tarr2018hack}. Timbral adjectives often have a correlation with the parameter setting for the equalizer. Some examples include,  \emph{add air}, \emph{make it warmer}, and \emph{make it less muddy}~\cite{spyridon2019audio}. Kulka~\cite{kulka1972equalization} associated adjectives such as \emph{warmth}, \emph{honk}, \emph{crunch}, and \emph{sibilance} with frequencies of 125, 500, 2000, and 8000~Hz respectively. For example, according to the Kulka rule, if the mix sounds honky, cut the region around 500~Hz. 

When clients such as instrumentalists and musical directors work with mixing engineers, they often use semantic descriptors to describe their goals. For example,\emph{ ``make the violin sound warmer''}~\cite{cartwright2013social}. It is the role of the mixing engineer to understand these descriptors. Popular semantic descriptors such as \emph{warm} and \emph{bright} are easily understood by the mixing engineer~\cite{bromham19brightness}. To expand the vocabulary of such descriptors, studies have also tried to create a thesaurus with synonyms and antonyms. For example, significant synonyms of \emph{boom} are \emph{boxy, dull, and fat} and significant antonyms of \emph{boom} are \emph{air, bright, and crisp}~\cite{spyridon2019audio}. However, the problem arises when individuals without training in audio production describe their creative goals~\cite{zheng2016socialfx}. They may have ideas that cannot be directly translated into a studio engineer's vocabulary.

\begin{table*}[t]
	\tabcolsep8.1pt
	\tbl{Four cross-validation folds from the dataset. The test words from each fold are presented in the table. For each fold, the training set consists of words that are not in the test set.\label{table:test-folds}}{
		\begin{tabular}{p{3.4cm}p{3.4cm}p{3.2cm}p{3.2cm}}
			\toprule
		\textbf{Fold 1} & \textbf{Fold 2} & \textbf{Fold 3} & \textbf{Fold 4} \\ \colrule
		\textbf{smooth,
		muffled,
		crisp,
		punch,
		clean,
		brittle,
		muddy,
		soothing,
		clear,
}		brassy,
		caring,
		mellow,
		throbbing,
		cooing,
		fluffy,
		good,
		excited,
		squeaking,
		punchy,
		funky,
		whispered,
		disgusting,
		beautiful,
		reserved,
		serene,
		thumpy,
		pleasurable,
		whispering,
		gentle,
		energetic,
		peace & 
\textbf{crunchy,
		woody,
		flat,
		metallic,
		dull,
		tinny,
		cold,
		booming,
		deep,
}		energizing,
		heart-warming,
		edgy,
		heavy,
		edge,
		strong,
		enchanting,
		cheerful,
		plodding,
		quiet,
		radiant,
		biting,
		brass,
		pleasing,
		light,
		taco,
		gruff,
		exciting,
		love,
		heat,
		techno,
		solemn & \textbf{sweet,
		warm,
		airy,
		full,
		boxy,
		bright,
		boom,
		fat,
		shrill,
}		calm,
		velvety,
		hard,
		rich,
		noisy,
		down,
		rumble,
		sloppy,
		relaxing,
		peaceful,
		romantic,
		low,
		hot,
		thunderous,
		frigid,
		happy,
		poor,
		cool,
		tense,
		jagged,
		forceful,
		aggressive & \textbf{sharp,
		big,
		dark,
		hollow,
		harsh,
		smooth,
		muffled,
		crisp,
		punch,
}		mournful,
		clarity,
		genius,
		bold,
		twangy,
		soft,
		splash,
		slow,
		wistful,
		brash,
		fancy,
		cute,
		rousing,
		loud,
		breezy,
		large,
		passionate,
		baseball,
		huge,
		icy,
		brassy,
		caring \\
 \botrule
	\end{tabular}}
\end{table*}

To address this issue of non-technical descriptors, Cartwright and Pardo~\cite{cartwright2013social} presented a dataset called SocialEQ, which is a web-based project that adopts crowd-sourcing to learn a vocabulary of audio descriptors. As it is crowd-sourced, the study focuses on aggregating a vocabulary to enable non-technical individuals to describe their sonic goals. Crowd-sourcing was also adopted to build the datasets for other effects like reverberation~\cite{seetharaman2014crowdsourcing} and dynamic range compression~\cite{zheng2016socialfx}. 

There is a growing interest in adopting natural language processing (NLP) methodologies to develop semantically-controlled audio effects~\cite{zacharakis2012analysis, williams2007perceptually, miranda1995artificial}. Stables~et~al.~\cite{stables2014safe} presented a system called Semantic Audio Feature Extraction (SAFE), which focused on extracting semantic descriptions for equalization from a digital audio workstation (DAW). Stasis~et~al.~\cite{stasis2016semantically} investigated the idea of mapping the descriptors to a reduced dimensionality space, to enable users to interact with the system in a more intuitive way. Chourdakis et al.~\cite{chourdakis2019tagging} explored tagging and retrieval of room impulse responses for reverberation. They adopted word embeddings to assign impulse responses to tags that match their short descriptions.

In this paper, we explore the novel idea of adopting word embeddings to automatically predict EQ settings. We present a methodology to translate words from a semantic vector space to a vector space representing the parameters of an equalizer. Word embeddings are representations of words that capture lexical semantics in language~\cite{bakarov2018survey}. An embedding layer is often used as the first layer in a neural network that performs NLP tasks such as machine translation, caption generation, and automatic speech recognition~\cite{goldberg2017neural}. Although word embeddings are commonly used to understand natural language, we investigate if they would be of any benefit to descriptors for EQ settings. We adopt this approach to translate words to predict values of a parametric equalizer. This way, the neural network has the ability to understand non-technical words and even descriptors that it has not seen before. This finding is significant because artists without training in audio production can express their creative goals directly to the AI-powered mixing engine. To our knowledge, this is the first study that investigates how EQ settings can be predicted for \emph{unseen} semantic descriptors. We demonstrate that the neural network is capable of learning a direct translation from the text domain to the EQ domain.



\section{METHODOLOGY}

\subsection{DATASET}
We adopted the SocialEQ dataset~\cite{cartwright2013social}, which crowd-sources semantic descriptors for EQ settings. In the raw format, each sample in the dataset contains a semantic descriptor, language of the descriptor, audio id, a consistency rating, and 40 values for EQ parameters. During the data collection, each participant was asked to enter a word in their preferred language. For example, \emph{warm} in English, \emph{claro} in Spanish, or \emph{grave} in Italian. Subsequently, they pick a sound file which will be modified by the EQ plugin. There were three sound files --- electric guitar, piano, and drums. Each sound file had a unique audio id.

After selecting a descriptive term and audio file, the participant was presented with 40 different modifications of the sound file made by different EQ settings. Suppose the user has selected \emph{warm}, they are asked to rate \emph{how warm that sound is}. Out of the 40 modifications, there are 15 repetitions to test for consistency. Consistency score was calculated using Pearson correlation between the ratings of the test and repeated examples. The system processes the ratings of the user and develops a relative boost/cut for 40 different frequency bands. Refer to Cartwright and Pardo~\cite{cartwright2013social} for more details on the dataset.

The dataset has 1595 samples in it. For simplicity, we considered only descriptors in English. The number of examples in English was 918. It is important to note that the dataset contained examples with different EQ parameter settings for the same word. Thus, the number of unique descriptors in English was 388. 

\subsection{TRAIN-TEST SPLIT}
\label{sec:train-test-split}
An important hypothesis we wanted to test in this paper is that a word embedding layer helps a model predict EQ parameter settings for semantic descriptors it has not seen before. Therefore, words in the test set should not appear in the training set. We adopted a four-fold cross-validation setup~\cite{forman2010apples} and the strategy is explained below.

We aggregated a list of semantic descriptors that are common in the audio mixing literature. We labelled these as High Quality (HQ) words. In order to avoid bias and objectively choose these words, we selected those that were already listed in table 4.8 in Ref.~\cite{spyridon2019audio}. Additionally, we included semantic descriptors that fell under the hierarchical ontology presented by Pearce~et~al.~\cite{pearce2016audio}. The list of HQ words are presented in bold in table \ref{table:test-folds}. There are 32 HQ words present in the SocialEQ dataset.

We also aggregated a list of words that were Highly-Rated (HR). HR words need be not semantically meaningful, but have a high consistency score in the dataset. Words that have a consistency score greater than 0.7 were selected as HR words. As these words have a high consistency score, the user strongly associated the semantic word with a particular EQ setting. Words in table \ref{table:test-folds} that are not formatted as bold text are HR words. Totally, 86 HR words were present in the SocialEQ dataset.

Each test fold contained 9 HQ words and 22 HR words. We ensured that every HQ and HR word was tested at least once. In the last test fold, there may be a few repetitions of words from the first test fold. There was no overlap between the training set and test set. The test set only contained words that were not present in the training set. Note that the network for each fold is trained as a separate experiment. In other words, the network is totally trained four times and tested four times on different folds and we report the average performance. 

As mentioned earlier, each word can have multiple EQ settings. Each setting is a separate example and can have different consistency scores. In the test set, we only included examples that had a consistency score of greater than 0.7. In the training set, we did not exclude any words based on the consistency score.

\subsection{WORD EMBEDDINGS}
A vocabulary consists of all the possible words that the neural network can understand. Generally, a word is converted into a one-hot encoded vector before passing into the neural network. For instance, in the SocialEQ dataset, there are 388 unique words, which means that the size of the vocabulary is 388. Therefore, the dimensions of the one-hot encoded vector are $1\times388$. Each position within the vector is assigned to a unique word. Thus, the respective position of the word is labelled as 1 and the remaining elements are 0. However, it is important to note that the Euclidean distance between any pair of words is equal. As each word is equidistant from each other, the neural network is not capable of handling words that are not present in the training set. For example, let us consider the semantic descriptor \emph{bright} and assume that it is present in the training set. Let us also assume that \emph{clear} and \emph{boom} are words in the test set. According to Spyridon~\cite{spyridon2019audio}, \emph{clear} is a synonym of \emph{bright} and \emph{boom} is an antonym of \emph{bright}. Thus, we expect similar EQ settings for \emph{clear} and \emph{bright}, but considerably different EQ settings \emph{boom} and \emph{bright}. However, the neural network cannot perceive this understanding unless it has seen all three words because each word is equidistant from each other. Furthermore, this issue becomes exaggerated if a non-technical user is utilising a semantic descriptor that is not common in the audio mixing literature.

A word embedding layer converts a one-hot encoded representation into a vector space of reduced dimensionality. Large vocabularies with millions of words can be reduced to a 300-dimensional vector representation~\cite{pennington2014glove}. The distances between words in the embedding space are governed by some form of semantic correlation. Examples include synonyms or two words frequently occurring together. There are different algorithms to train word embedding models. Some of them include Word2Vec~\cite{mikolov2013efficient}, GloVe~\cite{pennington2014glove}, ConceptNet~\cite{speer2017conceptnet}, and Dict2Vec~\cite{tissier2017dict2vec}. Each of these algorithms presents unique methods to train on large corpora of text such as Wikipedia. Effectively, they try to learn semantic relationships between words and represent them through an embedding vector.

For this study, we investigated four different embedding models --- GloVe-6B, GloVe-840B, Tok2Vec, and Dict2Vec. GloVe is an unsupervised learning algorithm developed to obtain vector representations for words~\cite{pennington2014glove}. GloVe-6B refers to the model that was trained on Wikipedia~2014 and Gigaword~5. It includes 6B tokens and a vocabulary size of 400k (B, M, k stand for Billion, Million, and thousand respectively). Moreover, GloVe-840B uses 840B tokens and a vocabulary size of 2.2M. It trains on the World Wide Web using Common Crawl, which is a larger corpus of text. Tok2Vec is a word embedding model provided by a company called spaCy~\cite{honnibal2020spacy}. We did not find the entire details regarding its implementation, but the model is publicly available and free to use. 

It is important to note that word embeddings are used for NLP tasks, which are designed to accept sentences. In our application, we are considering only one word, which is the semantic descriptor. As GloVe and Tok2Vec also focus on the ordering of words in sentences, we thought it is a good idea to consider another embedding model called Dict2Vec~\cite{tissier2017dict2vec}. Dict2Vec is an embedding model that uses lexical dictionaries. It builds new word pairs from dictionary entries so that semantically-related words are closer to each other in the embedding space~\cite{tissier2017dict2vec}. Similar to GloVe-6B, it was trained on the Wikipedia corpus.



%
%
%
%

\subsection{MACHINE LEARNING ARCHITECTURE}
\subsubsection{WORD EMBEDDING LAYER}
We evaluated four different pre-trained word embedding models in the study --- GloVe-6B, GloVe-840B, Tok2Vec, and Dict2Vec. All the models represent words with 300-D semantic vectors. This is convenient because we can adopt the same neural network architecture to compare different embeddings. Initially, a word is converted into a one-hot encoded representation. Subsequently, an embedding matrix converts this one-hot encoded representation into a 300-D semantic vector. Then, this vector is connected to hidden layers in the network. Note that the weights of the embedding matrix are frozen and the layer is not trainable. We did not consider setting this to trainable because of the limited data we have.


\subsubsection{HIDDEN LAYERS}

\begin{figure}[t]
	\centering
	\includegraphics[trim={0.4cm 0.4cm 0.4cm 0.2cm},width=\linewidth]{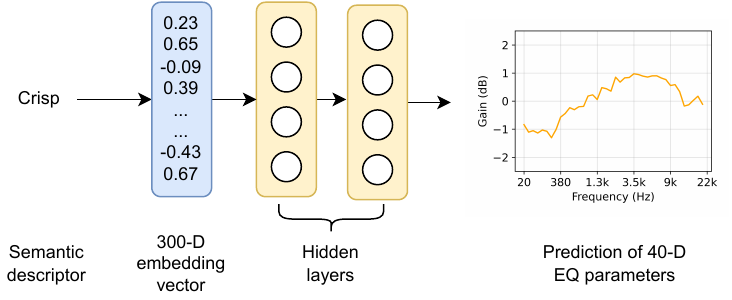}
	\caption{A schematic diagram of how the network learns a translation from semantic descriptors to EQ parameters.}
	\label{fig:plot-fold1and2}
\end{figure}

\begin{table}[t]
	\tabcolsep8.1pt
	\tbl{The neural network architecture.\label{table:architecture}}{%
		\begin{tabular}{@{}cccc@{}}\toprule
			\textbf{Layer type}& \textbf{Units} & \textbf{Activation} & \textbf{Output shape}\\ \colrule
			Embedding  &  -     & -          & 300          \\
			Dense      & 300   & ReLu       & 300          \\
			Dense      & 200   & ReLu       & 200          \\
			Dense      & 100    & ReLu       & 100           \\
			Dense      & 80    & ReLu       & 80           \\
			Dense      & 60    & ReLu       & 60           \\
			Dense      & 40    & Sigmoid     & 40          \\\botrule
	\end{tabular}}
	
\end{table}

The neural network aims to translate a representation of word embeddings to a prediction of equalizer parameters. Therefore, our network needs to be deep enough to learn the translation between two domains. Deeper networks apply the non-linear activation more number of times on the input and therefore have the advantage of learning more complex translations. However, it is important to note that our dataset is relatively small for our task.

All the layers in the neural network were fully connected layers. Table \ref{table:architecture} shows an overview of the architecture. After the embedding layer, we had a series of fully connected layers. The number of hidden units in these layers were 300, 200, 100, 80, and 60 respectively. Finally, it was connected to an output layer with 40 units. Excluding the final layer, all the hidden layers were fitted with ReLu activations and a dropout of 0.1. The output layer is explained in section \ref{sec:output-layer}. The code and trained models associated with this study can be found in this GitHub repository~(\href{https://github.com/satvik-venkatesh/word-eq}{https://github.com/satvik-venkatesh/word-eq}).

\subsubsection{NORMALISATION}
Traditional min-max normalisation by calculating the maximum and minimum in the training set was not appropriate for our dataset. This is because if there exists any outliers amongst the values in the test set, specific features may get magnified or diminished. Furthermore, as we are predicting values for 40 EQ bands, this issue becomes more crucial. Therefore, we fixed the minimum and maximum value for each EQ parameter to -4~dB and +4~dB respectively. In other words, the highest cut/boost within each EQ band was 4~dB. The values were linearly normalised to the range of 0 to 1. Hence, -4~dB would correspond to 0 and +4~dB would correspond to 1 in the output layer.

\subsubsection{OUTPUT LAYER AND LOSS FUNCTION}
\label{sec:output-layer}
The output layer of the network contained 40 neurons, with each of them predicting a value for one EQ band. As we normalised the data within the range of 0 to 1, we used sigmoid activation functions for the output neurons. As we are working with a regression problem, we adopted the mean absolute error loss function, which is commonly adopted by many studies. All EQ bands were given equal importance when averaging the error for the loss function. In future work, it would be interesting to weight the EQ bands based on perceptual frequency band weights. However, that is beyond the scope of this study.

The network was trained using stochastic gradient descent with an initial learning rate of 0.1. The learning rate was scaled by 0.96 after every 10,000 weight updates.



\section{Results}
\subsection{Error}
\begin{table}[t]
	\tabcolsep11.1pt
	\tbl{The error calculated across four folds. The smallest error is indicated in bold.\label{table:error}}{%
\begin{tabular}{cc}\toprule
	\textbf{Word Embedding} & \textbf{Error} \\ \colrule
	Tok2Vec                                       & \textbf{0.760 $\pm$ 0.055}                    \\
	Glove-840                                   & 0.770 $\pm$ 0.032                    \\
	Dict2Vec                                    & 0.792 $\pm$ 0.058                    \\
	Glove-6B                                    & 0.798 $\pm$ 0.046                   \\
	No Embedding                                & 0.836 $\pm$ 0.016                   \\ \botrule
\end{tabular}}	
\end{table}

\label{sec:results-error}
Table \ref{table:error} shows the mean absolute error for different embedding models calculated across four test folds. As we can see, Tok2Vec obtains the lowest error of 0.76, followed by GloVe-840 with an error of 0.77. GloVe-840 obtains an error lower than GloVe-6B, which conveys that it benefited from training on a larger corpus. Dict2Vec and GloVe-6B were trained on similar dataset sizes and the former obtained a better error. This suggests that the performance of Dict2Vec can be improved with training on a larger corpus of text.

The `No Embedding' model in table \ref{table:error} means that no word embedding layer was used in the neural network. This can be considered to be the baseline system. As this is the first study that investigates a translation from \emph{unseen} semantic descriptors to EQ settings, there are no state-of-the-art approaches for comparison. The input of the network was a direct one-hot encoded representation. All the neural networks with word embeddings performed better than the model without word embeddings. However, the difference was not huge. The best model was Tok2Vec with an error of 0.76 vs `No Embedding' with an error of 0.836. This is possibly due to two reasons. Firstly, error may not be the best metric for our task. For example, the semantic word \emph{warm} may have a boost of 1.2~dB at 260~Hz. But, the neural network may predict a boost at the adjacent EQ band, such as 317~Hz. Although, the error in this case is high, the EQ effect applied to the audio may still be semantically meaningful. Secondly, the test set contains many semantic descriptors that occur only once. These examples may be highly subjective to one individual, despite having a high consistency score. Therefore, in the next subsection, we evaluate the top two performing models using Partial Curve Mapping (PCM) \cite{witowski2012parameter}, which is a method to quantify the similarity between two curves. For instance, this technique is generally adopted to analyse similarities between hysteresis curves pertaining to a magnetic field. Although this technique may not be ideal for our task, it would give us a better understanding of our model's performance compared to mean absolute error.

\subsection{PARTIAL CURVE MAPPING}
\label{sec:pcm}
\begin{figure}[t]
\centering
\includegraphics[trim={0.4cm 0.4cm 0.4cm 0.2cm},clip,width = 0.8\linewidth]{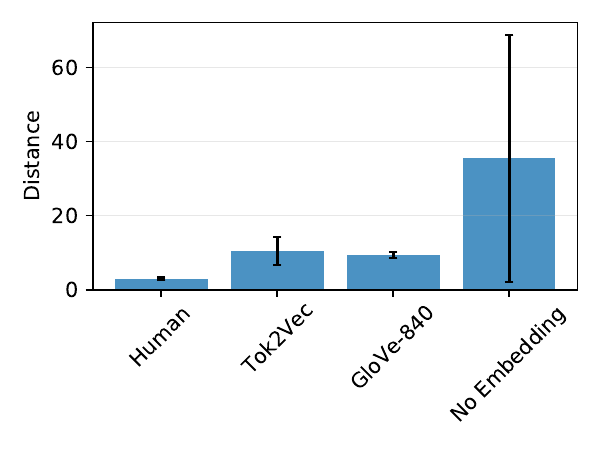}
\caption{Distances obtained by different models calculated by using PCM. An ideal algorithm would have a distance of zero.}
\label{fig:pcm}
\end{figure}

\begin{figure*}[t]
	\centering
	\includegraphics[trim={0.4cm 1.8cm 0.4cm 0.2cm},clip, width=0.70\linewidth]{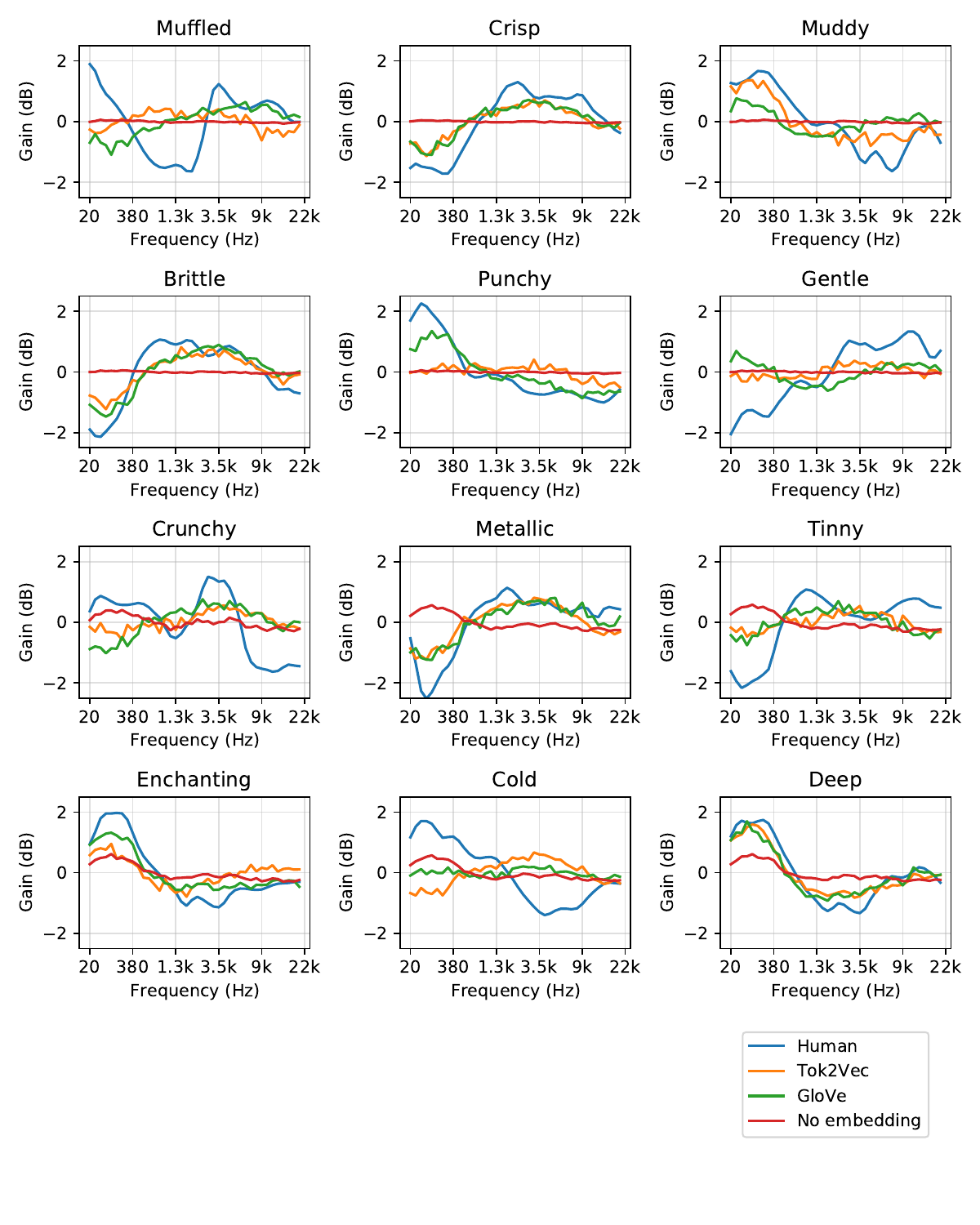}
	\caption{Plots of EQ parameters for words in test folds 1 and 2. Note that each word in the test set does not occur in the training set. The first two rows occur in fold 1 and the last two rows occur in fold 2. The human label plotted for a semantic word was the EQ settings with the highest consistency score in the dataset.}
	\label{fig:plot-fold1and2}
\end{figure*}

\begin{figure*}
	\centering
	\includegraphics[trim={0.4cm 1.8cm 0.4cm 0.2cm},clip, width=0.70\linewidth]{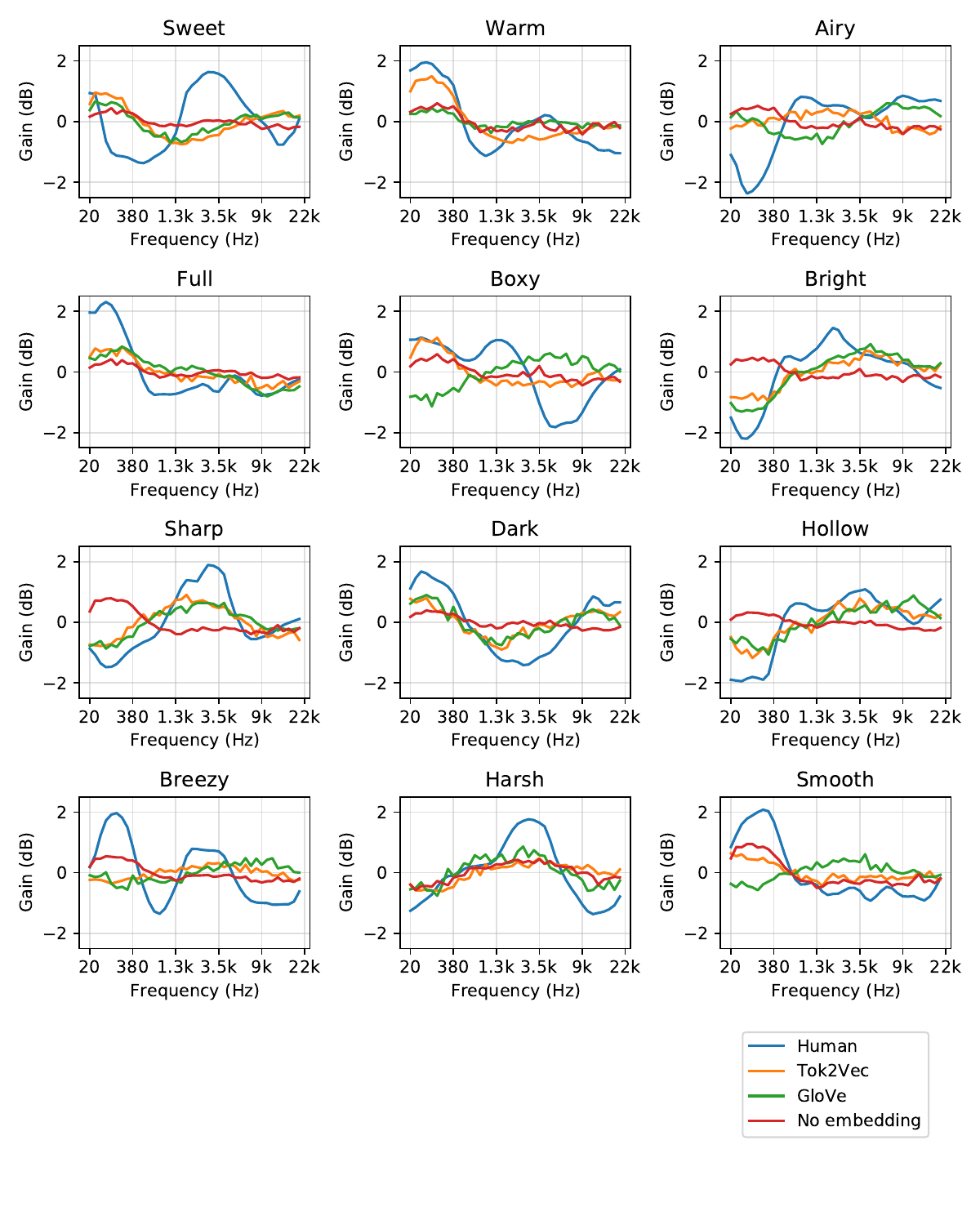}
	\caption{Plots of EQ parameters for words in test folds 3 and 4. The first two rows occur in fold 3 and the last two rows occur in fold 4.}
	\label{fig:plot-fold3and4}
\end{figure*}

In this section, we evaluate the models using PCM. PCM was implemented using this Python package \cite{jekel2019similarity}. We also compare our model to human labels. As mentioned earlier, each semantic descriptor had multiple EQ settings in the dataset. To calculate the error in human labels, we considered the mean of the different EQ settings as the ground truth. However, words that occur only once in the dataset would not have an error associated with it. These words would artificially reduce the average error. Hence, we only included words that occur at least twice in the dataset. Figure \ref{fig:pcm} shows the distances for different models. An ideal algorithm would obtain a distance of zero. Human labels obtain the smallest distance of 2.9, which is an expected observation. GloVe and Tok2Vec obtain similar distances with the former performing slightly better. The distances were 9.3 and 10.5 respectively. Note that for this experiment, we only considered words that occur at least twice, which is different from results presented in section \ref{sec:results-error}. The mean distance of the model with no embeddings was 35.4, which was considerably higher. In addition, there was a much larger standard deviation for this model, which suggests that it was randomly guessing.

\begin{figure*}[t]
	\centering
	\includegraphics[trim={0.4cm 1.8cm 0.4cm 0.2cm},clip, width=0.70\linewidth]{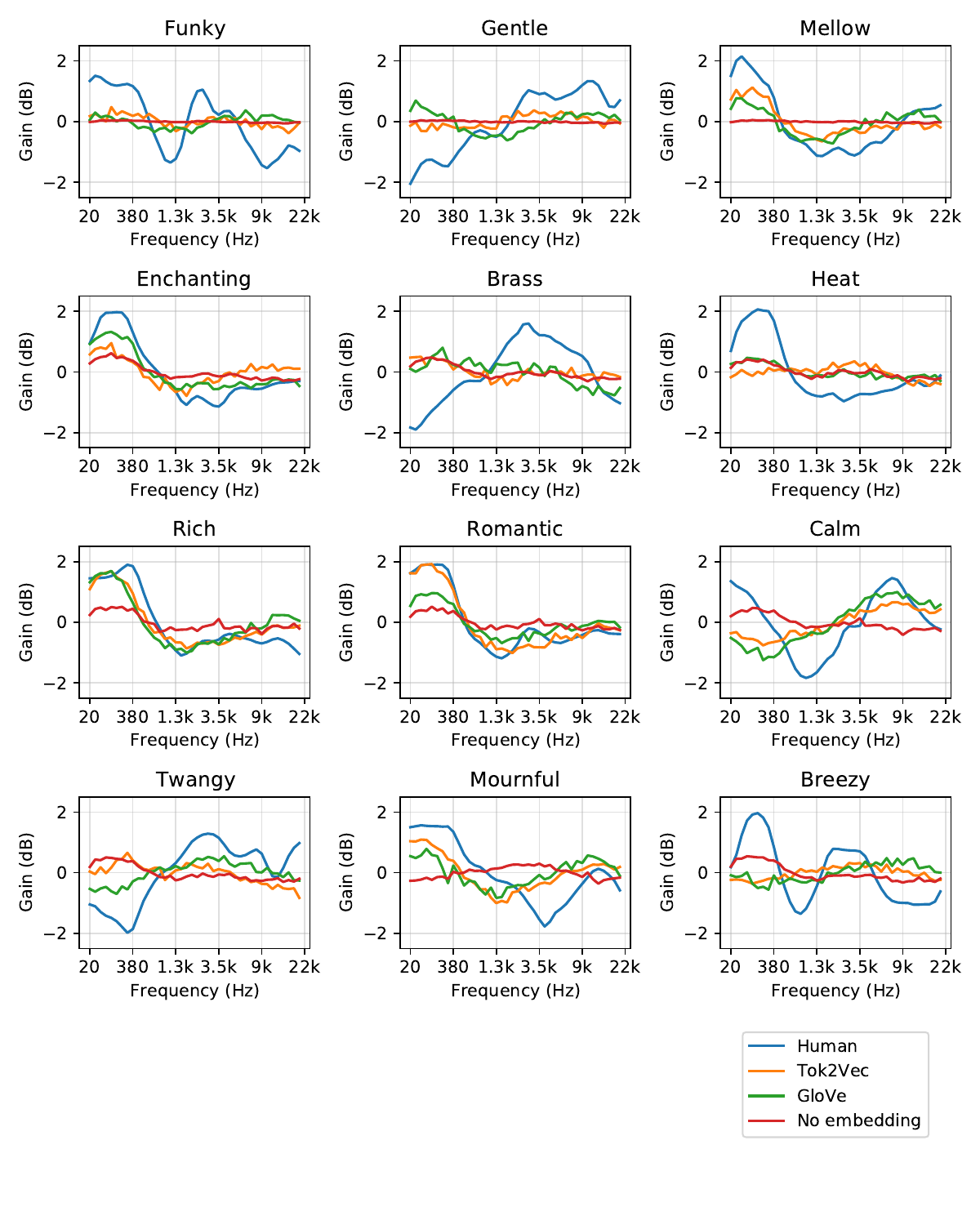}
	\caption{Plots of EQ parameters for highly-rated (HR) words as explained in section \ref{sec:train-test-split}. These are non-technical words that may be highly subjective to a user.}
	\label{fig:hr-plots}
\end{figure*}

\subsection{PLOTS OF EQ PARAMETERS}
\label{sec:results-plots}
In this section, we perform an error analysis of predictions made by the machine learning models. We look at individual test words to investigate if the neural network is actually learning semantic meanings. We predominantly look at HQ words as they are common in the audio mixing literature and would be more intuitive to evaluate. In figure \ref{fig:plot-fold1and2} and \ref{fig:plot-fold3and4}, we plot the EQ settings of human labels alongside the predictions of Tok2Vec, Glove-840B, and `no embedding'. As the literature does not comprise an `ideal' metric for our task of predicting EQ parameters, we plotted graphs and actually visualised the predictions of the algorithms. Figure \ref{fig:plot-fold1and2} plots the graphs for words selected from the test folds 1 and 2. Figure \ref{fig:plot-fold3and4} plots the graphs for words selected from the test folds 3 and 4. Note that for each word in the test folds, the neural network has not encountered the word in the training set. The human label chosen for each semantic word in the plots was the EQ setting with the highest consistency score in the dataset.

In figure \ref{fig:plot-fold1and2}, human labels for \emph{muffled} had boosts at 20~Hz and 3.5~kHz. For Tok2Vec and GloVe, we saw slight boosts in the mid-range and high-range respectively, which may convey that the neural networks did not interpret this word correctly. We also observed that the predictions made by Tok2Vec and GloVe are considerably different from each other. This can be due to two reasons --- (1) Tok2Vec and GloVe are different algorithms and therefore, learn different semantic meanings from text (2) there may be a higher degree of randomness in their predictions because the the embeddings are trained only on natural text from the Word Wide Web, which is different from EQ descriptors. Hence, the neural network would require more training examples containing EQ descriptors. The network with `no embedding' was basically a flat curve for all the words in the first fold. For \emph{crisp}, interestingly, the predictions of Tok2Vec and GloVe did follow a similar pattern as the human labels. In the human labels, we saw boosts at 2.1k and 9k. For GloVe and Tok2Vec, we saw a gradual boost at 3k, which lifts the high-range of the frequency spectrum. Some semantic synonyms of \emph{crisp} present in the training set for this respective fold include \emph{bright}, \emph{harsh}, \emph{hollow}, and \emph{sharp}. This means that the word embedding has delineated a relationship between the semantic word and EQ predictions. Again, as mentioned earlier, we did not observe a meaningful pattern in the neural network with `no embedding' because the curves were flat.

\emph{Muddy} had a gradual boost from 200 to 380~Hz in the human labels. Tok2Vec follows a very similar pattern in its prediction by boosting the lows and cutting the highs. GloVe's prediction has slightly boosted lows and highs, which is not convincing for the semantic word \emph{muddy}. Some semantic synonyms in the training set include \emph{boom}, \emph{muddled}, \emph{dark}, \emph{dull}, and \emph{fat}. The next test word, \emph{brittle} was well-understood by both Tok2Vec and GloVe. There was considerable overlap with the human labels. The synonyms for \emph{brittle} in the training set would be similar to those listed for \emph{crisp}. \emph{Punchy} was understood GloVe, but not by Tok2Vec. \emph{Gentle} was not understood by both embedding models (please refer to table \ref{table:test-folds} for more semantic synonyms. If fold 1 is selected as the test set, folds 2, 3, and 4 are included in the training set).

\emph{Crunchy} had boosts in the low and high frequency range in the human labels. We observe a boost for GloVe and Tok2Vec in the high range. The `no embedding' model has a boost in the low range. However, if you observe, it has made the same prediction for all the test words in the second fold. GloVe and Tok2Vec and correctly understood the semantic descriptor \emph{metallic} and have significant overlap with human labels. For \emph{tinny}, human labels have boosts at 1.3k and 9k. Whereas, the neural networks with embeddings have a gradual boost around 3k. We are not certain if these predictions would have a \emph{tinny} effect. For test words \emph{enchanting} and \emph{deep}, we observed a noticeable overlap with human labels. However, for \emph{cold}, it seems as though GloVe and Tok2Vec predicted the antonym.

In test fold 3, \emph{sweet} was not understood by the networks at all. For \emph{warm}, Tok2Vec has a noticeable overlap with the human labels because both have a boost of approximately 2~dB in the low frequency range. \emph{Airy} was partially convincing because GloVe recognised a boost at 9~kHz. Although, the networks have boosted the lows for \emph{full}, it seems like a random guess as the prediction significantly overlaps with the one made by `no embedding'. The predictions made by the networks for \emph{boxy} were not convincing. \emph{Bright} seemed plausible with Tok2Vec and GloVe boosting the high frequency range.

In test fold 4, we saw reasonable overlap for \emph{sharp}, \emph{dark}, \emph{hollow}, and \emph{harsh}. We did not observe a reasonable pattern for \emph{breezy} and \emph{smooth}. Interestingly, the network with `no embedding' predicted the EQ settings for \emph{harsh} correctly. This is a chance occurrence because it predicted a standard template of settings for all the other words. 

In figure \ref{fig:hr-plots}, we analyse the predictions on non-technical words. These non-technical words are the same as the HR words explained in section \ref{table:test-folds}. Although these words may have a high consistency score in the SocialFX dataset, they may be highly subjective to the user. However, we compared the predictions of GloVe and Tok2Vec to the human labels. There was considerable overlap for \emph{mellow}, \emph{enchanting}, \emph{rich}, and \emph{romantic}. For \emph{mournful} and \emph{calm}, there were similar patterns between the predictions of the word embedding models and human labels. However, for \emph{heat} and \emph{brass}, the word embedding models did not predict a relevant pattern. Although the training set contained semantically similar words like \emph{warm} and \emph{brassy}, the embeddings did not perceive these similarities. This conveys that the algorithms to learn word embeddings can be further optimised for EQ mixing.

\section{DISCUSSION}
In the previous section, we show that a word embedding layer is helpful for automatic mixing. We analysed the error of models in section \ref{sec:results-error}. All the models with an embedding layer obtained lower errors than the one without an embedding layer. We further analysed the performance of GloVe-840B and Tok2Vec by using partial curve mapping. The mean distances obtained by human labels, Tok2Vec, GloVe, and `no embedding' were 2.9, 10.5, 9.3, and 35.4 respectively. This objectively demonstrates that the embedding models perform better than models without an embedding layer, but not as good as human labels. 

In section \ref{sec:results-plots}, we conducted an error analysis of predictions made by GloVe and Tok2Vec. We observed that the machine learning models were able to understand semantic descriptors that they had not encountered before. This is a promising step towards understanding semantic descriptors from non-technical users. It is important to note the word embedding layers used in the networks were trained on corpora of written text. This concludes that there exists some common ground for semantic relationships between words in written text and for those adopted in EQ mixing. 

Considering the fact that we have adopted such a small training dataset, this performance is reasonable. The SocialFX dataset comprises only 388 unique English words. Additionally, many of the high-quality and highly rated words were used for testing in each fold. As our study has demonstrated that word embeddings are helpful for automatic EQ mixing, we hope to encourage researchers the build larger datasets with semantic descriptors. In the literature, another dataset called SAFE~\cite{stables2014safe} focused on extracting semantic descriptions for equalization from a DAW. We were unable to include the dataset within this study for two reasons. Firstly, as these are extracted directly from the DAW without post-processing, some labels can be noisy. Although the dataset contains many examples with meaningful descriptors, some words are randomly typed letters such as `xy', which have no semantic meaning. Perhaps, this noise may not matter when training the network with large-scale data. The second reason is that both datasets use different EQ plugins. The SocialFX dataset uses a 40 band EQ, whereas the SAFE dataset uses a five band EQ. We are not certain if additional noise would be induced in mapping one EQ domain to the other.

In this study, we analysed the performance of the machine learning model using objective metrics. However, it is important to perform listening tests with human participants to obtain subjective evaluations of the system. We need to investigate if users are satisfied with the way the machine learning model understands their semantic descriptors. After aggregating a larger dataset for this task, this could be a potential future pathway. 











\section{CONCLUSION}
In this paper, we demonstrated the feasibility of adopting word embeddings for automatic EQ mixing. We showed that the word embedding layer is capable of providing relationships between semantic descriptors, which assists in predicting EQ parameters. Using this technique, the machine learning model can predict EQ settings for words it has not seen before. This is a step towards bridging the gap between artists explaining their creative goals and mixing engineers understanding them.

In this study, we looked at EQ parameters as a separate entity. This may not be ideal in some scenarios. For example, the EQ settings for \emph{``make the vocals sound brighter''} maybe different from \emph{``make the drums sound brighter''}. Moreover, the number of EQ bands predicted was 40. This number is large for a network that performs regression. Future research could explore how the neural network architecture can be optimised and regularised better. Furthermore, it may be interesting to augment the size of training sets by adopting well-known synonyms and antonyms in the mixing engineer's vocabulary.

For some words, Tok2Vec captured relationships, but GloVe did not and vice versa. For example, GloVe captured the meaning of \emph{punchy} as shown in figure~\ref{fig:plot-fold1and2} and Tok2Vec captured the meaning of \emph{warm} as shown in figure~\ref{fig:plot-fold3and4}. This may be simply because there is limited data in the training set. Otherwise, different embedding models may capture different aspects of semantic relationships. Therefore, an ensemble of different embedding models will improve performance in this case. Furthermore, in our study, we discarded non-English words for simplicity. Word embedding models such as ConceptNet \cite{speer2017conceptnet} use a knowledge graph to connect words from different languages. This may be an interesting avenue to explore.

%
%
%
%
%
%
%
%
%
%
%
%


\section{ACKNOWLEDGMENT}
This paper is supported by EPSRC Grant EP/S026991/1 RadioMe: Real-time Radio Remixing. 

\bibliography{aes2e-sample}

\begin{thebibliography}{10}
\newcommand{\enquote}[1]{``#1''}
\providecommand{\url}[1]{\texttt{#1}}
\providecommand{\urlprefix}{URL }
\expandafter\ifx\csname urlstyle\endcsname\relax
  \providecommand{\doi}[1]{[Online]. Available: \discretionary{}{}{}#1}\else
  \providecommand{\doi}{doi:\discretionary{}{}{}\begingroup
  \urlstyle{rm}\Url}\fi

\bibitem{wilmering2020history}
T.~Wilmering, D.~Moffat, A.~Milo, M.~B. Sandler, \enquote{A history of audio
  effects,} \emph{Applied Sciences}, vol.~10, no.~3, p. 791 (2020 Jan.),
  \doi{10.3390/app10030791}.

\bibitem{ramirez2021deep}
M.~M. Ramirez, D.~Stoller, D.~Moffat, \enquote{A Deep Learning Approach to
  Intelligent Drum Mixing with the Wave-U-Net,} \emph{Journal of the Audio
  Engineering Society}, vol.~69, no.~3, pp. 142--151 (2021).

\bibitem{de2013knowledge}
B.~De~Man, J.~D. Reiss, \enquote{A knowledge-engineered autonomous mixing
  system,} presented at the \emph{135th Convention of the Audio Engineering
  Society} (2013 Oct.).

\bibitem{de2019intelligentmusic}
B.~{De Man}, R.~Stables, J.~D. Reiss, \emph{Intelligent Music Production}
  (Routledge) (2019 Oct.).

\bibitem{moffat2018towards}
D.~Moffat, F.~Thalmann, M.~B. Sandler, \enquote{Towards a semantic web
  representation and application of audio mixing rules,} presented at the
  \emph{4th Workshop on Intelligent Music Production, Huddersfield, UK} (2018
  Sep.).

\bibitem{perez2009automatic}
E.~Perez-Gonzalez, J.~Reiss, \enquote{Automatic gain and fader control for live
  mixing,} presented at the \emph{IEEE Workshop on Applications of Signal
  Processing to Audio and Acoustics}, pp. 1--4 (2009 Oct.),
  \doi{10.1109/ASPAA.2009.5346498}.

\bibitem{moffat2019machine}
D.~Moffat, M.~Sandler, \enquote{Machine learning multitrack gain mixing of
  drums,} presented at the \emph{147th Convention of the Audio Engineering
  Society} (2019 Oct.).

\bibitem{chourdakis2017machine}
E.~T. Chourdakis, J.~D. Reiss, \enquote{A machine-learning approach to
  application of intelligent artificial reverberation,} \emph{Journal of the
  Audio Engineering Society}, vol.~65, no. 1/2, pp. 56--65 (2017 Feb.),
  \doi{10.17743/jaes.2016.0069}.

\bibitem{moffat2019approaches}
D.~Moffat, M.~B. Sandler, \enquote{Approaches in intelligent music production,}
  presented at the \emph{Arts}, vol.~8, p. 125 (2019 Sep.),
  \doi{10.3390/arts8040125}.

\bibitem{tarr2018hack}
E.~Tarr, \emph{Hack Audio: An Introduction to Computer Programming and Digital
  Signal Processing in MATLAB} (Routledge) (2018).

\bibitem{spyridon2019audio}
S.~Spyridon, \emph{Audio equalisation using natural language}, Ph.D. thesis,
  Birmingham City University (2019 Jul.).

\bibitem{kulka1972equalization}
L.~d.~G. Kulka, \enquote{Equalization-the highest, most sustained expression of
  the recordist’s heart,} \emph{Recording Engineer/Producer}, vol.~3, pp.
  17--24 (1972 Nov.).

\bibitem{cartwright2013social}
M.~B. Cartwright, B.~Pardo, \enquote{Social-EQ: Crowdsourcing an Equalization
  Descriptor Map.} presented at the \emph{14th International Society for Music
  Information Retrieval (ISMIR) Conference}, pp. 395--400 (2013 Nov.).

\bibitem{bromham19brightness}
G.~Bromham, D.~Moffat, M.~Barthet, A.~Danielsen, G.~Fazekas, \enquote{The
  Impact of Audio Effects Processing on the Perception of Brightness and
  Warmth,} presented at the \emph{ACM Audio Mostly Conference} (2019 Sep.).

\bibitem{zheng2016socialfx}
T.~Zheng, P.~Seetharaman, B.~Pardo, \enquote{Socialfx: Studying a crowdsourced
  folksonomy of audio effects terms,} presented at the \emph{24th ACM
  international conference on Multimedia}, pp. 182--186 (2016 Oct.).

\bibitem{seetharaman2014crowdsourcing}
P.~Seetharaman, B.~Pardo, \enquote{Crowdsourcing a reverberation descriptor
  map,} presented at the \emph{22nd ACM international conference on
  Multimedia}, pp. 587--596 (2014 Nov.).

\bibitem{zacharakis2012analysis}
A.~Zacharakis, K.~Pastiadis, J.~D. Reiss, G.~Papadelis, \enquote{Analysis of
  musical timbre semantics through metric and non-metric data reduction
  techniques,} presented at the \emph{12th International Conference on Music
  Perception and Cognition (ICMPC12)}, pp. 1177--1182 (2012 Jul.).

\bibitem{williams2007perceptually}
D.~Williams, T.~Brookes, \enquote{Perceptually-motivated audio morphing:
  Brightness,} presented at the \emph{Audio Engineering Society Convention 122}
  (2007).

\bibitem{miranda1995artificial}
E.~R. Miranda, \enquote{An artificial intelligence approach to sound design,}
  \emph{Computer Music Journal}, vol.~19, no.~2, pp. 59--75 (1995 Jun.),
  \doi{10.2307/3680600}.

\bibitem{stables2014safe}
R.~Stables, S.~Enderby, B.~{De Man}, G.~Fazekas, J.~D. Reiss, \enquote{{SAFE}:
  A System for Extraction and Retrieval of Semantic Audio Descriptors,}
  presented at the \emph{15th International Society for Music Information
  Retrieval (ISMIR) Conference} (2014 Oct.), late Breaking/Demo.

\bibitem{stasis2016semantically}
S.~Stasis, R.~Stables, J.~Hockman, \enquote{Semantically controlled adaptive
  equalisation in reduced dimensionality parameter space,} \emph{Applied
  Sciences}, vol.~6, no.~4, p. 116 (2016 Apr.), \doi{10.3390/app6040116}.

\bibitem{chourdakis2019tagging}
E.~T. Chourdakis, J.~D. Reiss, \enquote{Tagging and retrieval of room impulse
  responses using semantic word vectors and perceptual measures of
  reverberation,} presented at the \emph{146th Convention of the Audio
  Engineering Society} (2019 Mar.).

\bibitem{bakarov2018survey}
A.~Bakarov, \enquote{A survey of word embeddings evaluation methods,}
  \emph{arXiv preprint arXiv:1801.09536} (2018 Jan.).

\bibitem{goldberg2017neural}
Y.~Goldberg, \enquote{Neural network methods for natural language processing,}
  \emph{Synthesis lectures on human language technologies}, vol.~10, no.~1, pp.
  1--309 (2017 Apr.).

\bibitem{forman2010apples}
G.~Forman, M.~Scholz, \enquote{Apples-to-apples in cross-validation studies:
  pitfalls in classifier performance measurement,} \emph{ACM SIGKKD
  Explorations Newsletter}, vol.~12, no.~1, pp. 49--57 (2010 Jun.),
  \doi{10.1145/1882471.1882479}.

\bibitem{pearce2016audio}
A.~Pearce, T.~Brookes, R.~Mason, \enquote{{Audio Commons}: Hierarchical
  ontology of timbral semantic descriptors,} Tech. rep. (2016).

\bibitem{pennington2014glove}
J.~Pennington, R.~Socher, C.~D. Manning, \enquote{Glove: Global vectors for
  word representation,} presented at the \emph{Conference on empirical methods
  in natural language processing (EMNLP)}, pp. 1532--1543 (2014 Oct.),
  \doi{10.3115/v1/D14-1162}.

\bibitem{mikolov2013efficient}
T.~Mikolov, K.~Chen, G.~Corrado, J.~Dean, \enquote{Efficient estimation of word
  representations in vector space,} \emph{arXiv preprint arXiv:1301.3781}
  (2013).

\bibitem{speer2017conceptnet}
R.~Speer, J.~Lowry-Duda, \enquote{ConceptNet at SemEval-2017 Task 2: Extending
  Word Embeddings with Multilingual Relational Knowledge,} presented at the
  \emph{11th International Workshop on Semantic Evaluation (SemEval-2017)}, pp.
  85--89 (2017 Aug.), \doi{10.18653/v1/S17-2008}.

\bibitem{tissier2017dict2vec}
J.~Tissier, C.~Gravier, A.~Habrard, \enquote{Dict2vec: Learning Word Embeddings
  using Lexical Dictionaries,} presented at the \emph{Conference on Empirical
  Methods in Natural Language Processing}, pp. 254--263 (2017 Sep.),
  \doi{10.18653/v1/D17-1024}.

\bibitem{honnibal2020spacy}
M.~Honnibal, I.~Montani, S.~{Van Landeghem}, A.~Boyd, \enquote{{spaCy}:
  Industrial-strength Natural Language Processing in Python,}  (2020 Nov.),
  \doi{10.5281/zenodo.1212303}.

\bibitem{witowski2012parameter}
K.~Witowski, N.~Stander, \enquote{Parameter identification of hysteretic models
  using partial curve mapping,} presented at the \emph{12th AIAA Aviation
  Technology, Integration, and Operations (ATIO) Conference and 14th AIAA/ISSMO
  Multidisciplinary Analysis and Optimization Conference}, p. 5580 (2012 Sep.),
  \doi{10.2514/6.2012-5580}.

\bibitem{jekel2019similarity}
C.~F. Jekel, G.~Venter, M.~P. Venter, N.~Stander, R.~T. Haftka,
  \enquote{{Similarity measures for identifying material parameters from
  hysteresis loops using inverse analysis},} \emph{International Journal of
  Material Forming} (2019 Jul.), \doi{10.1007/s12289-018-1421-8}.

\end{thebibliography}
\bibliographystyle{aes2e}

\biography{Satvik Venkatesh}{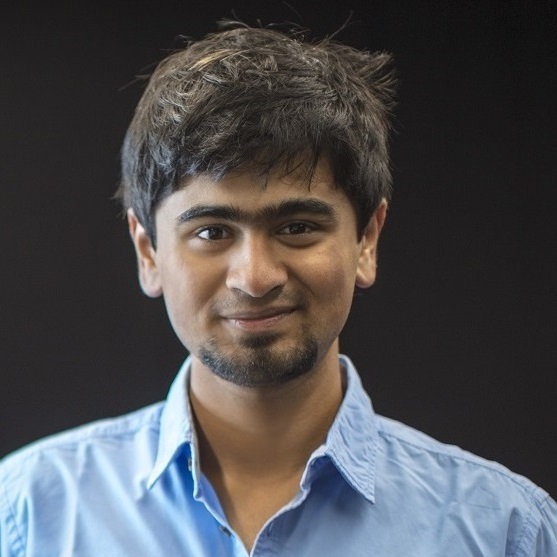}{Satvik Venkatesh holds a Bachelor of Technology in Information and Communication Technology from SASTRA University, India and a Master of Research in Computer Music from the University of Plymouth, UK. He currently is pursuing a PhD on the topic of audio segmentation and intelligent mixing for live radio broadcast. His research interests include Deep Learning, Brain-Computer Interfaces, and Unconventional Computing for music. Satvik is also an accomplished musician and performer.}

\biography{David Moffat}{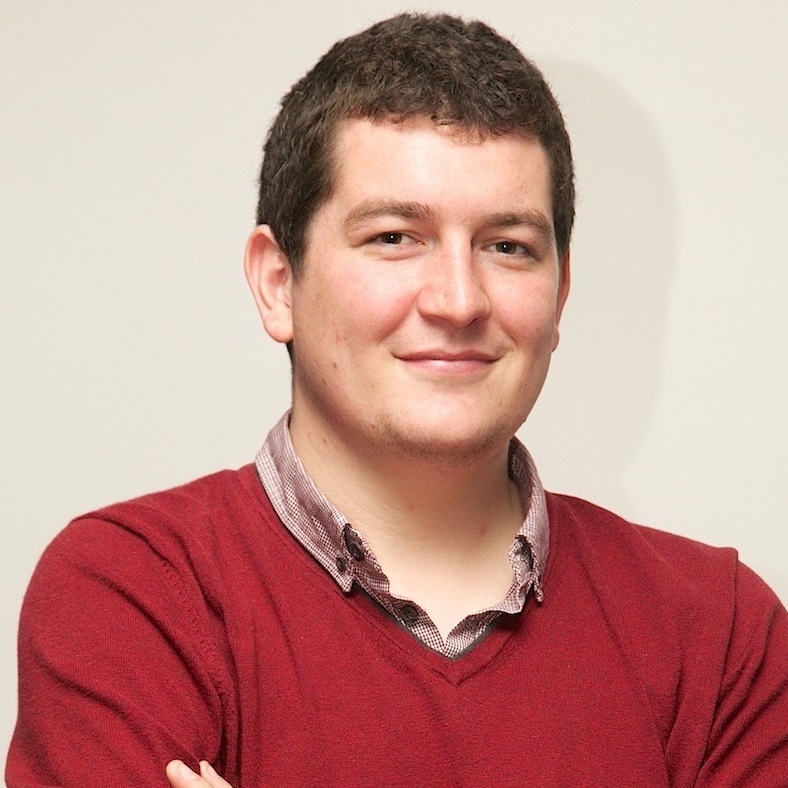}{David Moffat is an applied Artificial Intelligence (AI) researcher at Plymouth Marine Laboratory. Previously, he was a Lecturer in Sound and Music Computing at the University of Plymouth. He received his PhD from Queen Mary University in sound synthesis, machine learning, and subjective evaluation. His primary research interests are in the field of intelligent and assistive mixing and audio production through the implementation of semantic tools and AI.}

\biography{Eduardo Reck Miranda}{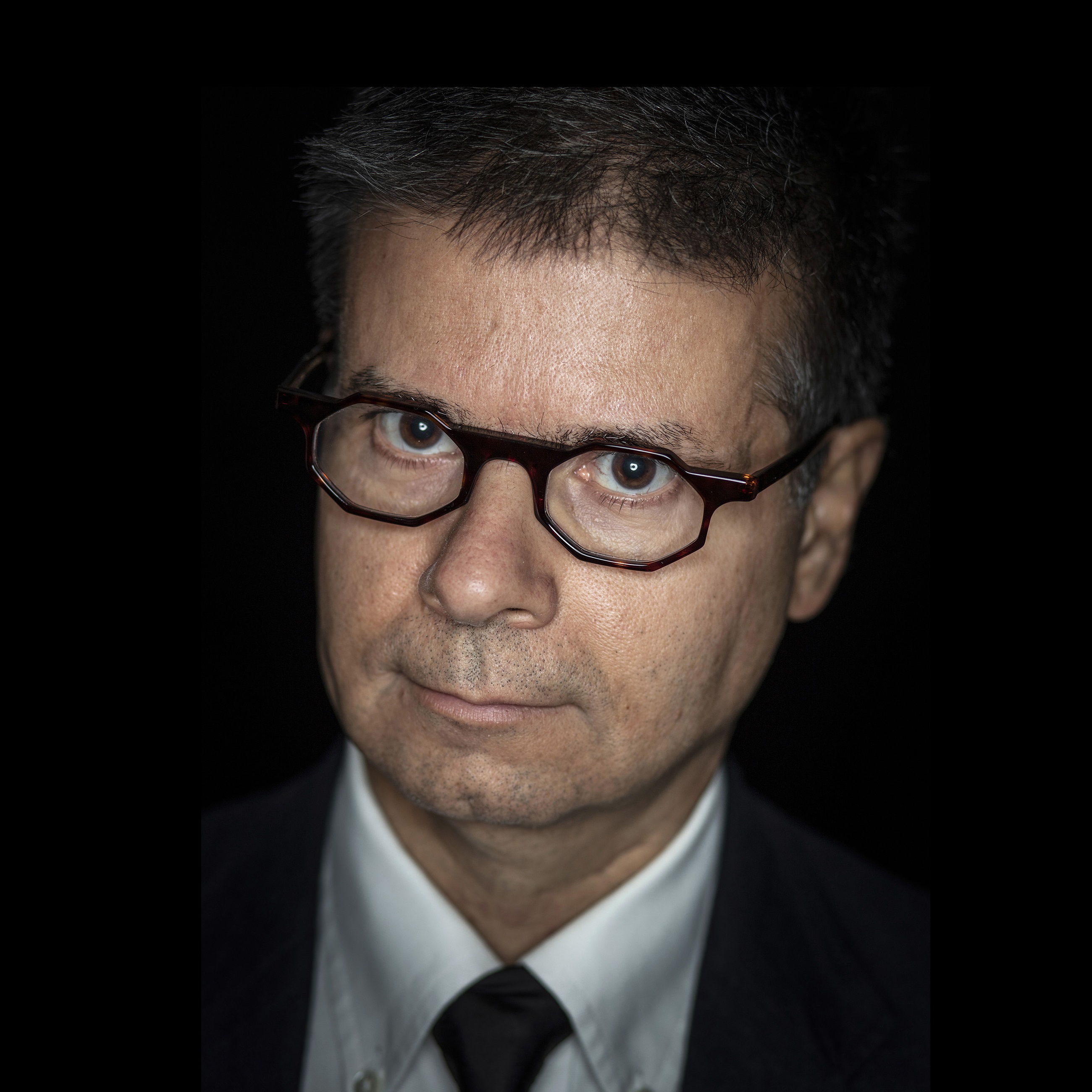}{Eduardo Reck Miranda is a composer and AI scientist working at the crossroads of biology and music. He received a PhD on the topic of musical composition with AI from the University of Edinburgh. Currently, he is a Professor in Computer Music at the University of Plymouth, where he leads the Interdisciplinary Centre for Computer Music Research, which is pioneering the fields of Music Neurotechnology and the development of biological and quantum computing for music.}
\end{document}